\documentclass[%
 reprint,
 amsmath,amssymb,
 aps,
]{revtex4-2}

\usepackage{graphicx}% Include figure files
\usepackage{color}% Align table columns on decimal point
\usepackage{dcolumn}% Align table columns on decimal point
\usepackage{bm}% bold math
\newcommand{\rev}[1]{\textcolor{black}{#1}}

\begin{document}

\title{Shear-Enhanced Elasticity In The Cubic Blue Phase I}

\author{Shuji Fujii}
\affiliation{Department of Food \& Life Sciences, Toyo University, Tokyo 112-0001, Japan}
\email[Email: ]{fujii034@toyo.jp}
\altaffiliation[Alternative address: ]{Faculty of Engineering, Division of Applied Physics, Hokkaido University, N13W8, Sapporo, Hokkaido 060-8628, Japan}

\author{Oliver Henrich}
\affiliation{Department of Physics, SUPA, University of Strathclyde, Glasgow G4 0NG, Scotland, UK}
\email[Email: ]{oliver.henrich@strath.ac.uk}

\date{\today}

\begin{abstract}
We present results of the linear and non-linear rheology of cubic blue phase I (BPI).
The elasticity of BPI is dominated by double-twist cylinders assembled in a body-centered cubic lattice, 
which can be specified by disclination lines.
\rev{We find that the elasticity of BPI is enhanced by an order of magnitude by applying pre-shear.}
The shear-enhanced elasticity is attributed to a rearrangement of the disclination lines 
that are arrested in a metastable state.
Our results are relevant for the understanding of the dynamics of disclinations in cubic blue phases.
\end{abstract}

%\keywords{Suggested keywords}%Use showkeys class option if keyword
                              %display desired
\maketitle

%\tableofcontents

%%%%%%%%%%%%%%%%%%%%%%
\section{Introduction}
%%%%%%%%%%%%%%%%%%%%%% 

Defects play a crucial role for the rheological properties of soft condensed matter, which shows a pronounced propensity to 
self-organize into mesoscale structures.
Not only the rheology of cholesteric liquid crystalline phases, but also that of the nematic and smectic phase is 
particularly strongly affected by the dynamics of intrinsic defects~\cite{HK,Basappa,MRMOS,MABK2,MAK,FKIL,LCIKK,DND}. 
Previous studies revealed that the elasticity of liquid crystals is mediated by their defects and that their viscosity 
is dominated by a creation-annihilation mechanism of defects under shear flow~\cite{LCIKK,Basappa,FKIL}.
Defects even affect non-equilibrium phase transitions in the thermotropic smectic phase~\cite{DND,FIKL2010,FKL2014,MNG,PAR,BLK}.
As defects influence all kinds of rheological phenomena in liquid crystals, it is therefore mandatory to study
 their dynamics in order to obtain a detailed understanding of fundamental aspects of structural rheology and 
topological defect networks.

In this study, we focus on the defect-mediated rheology of cholesteric blue phase I (BPI) with the goal to investigate the origin of its viscoelastic behavior.
BPI exists in a narrow temperature range
as one of the frustrated, intermediate phases between the cholesteric and the isotropic phase ~\cite{G56,S69,MSA1981,Crooker1989,Crooker,dGP}. 
In the presence of chiral constituents, the cholesteric host phase forms double twist cylinders (DTCs) as
its best compromise between minimizing its total free energy and accommodating 
more twist at the same time. This can be intuitively realized as in DTCs the director field rotates along all directions perpendicular to its axes as opposed to only one in a simple cholesteric phase. This rotation is usually decomposed into two rotations along the Cartesian axes perpendicular to the axis of the DTC, hence the naming `double-twist cylinder'. The DTCs arrange in regular, periodic arrays
with characteristic length scales in the range of several hundreds of nanometres. 
Importantly, DTCs cannot be formed without 
creating defect regions between them~\cite{MSA1981}.
This frustration gives rise to a network of topological $-1/2$ disclination lines, which is
responsible for many physical properties of BPs, including their rheology and characteristic behavior under phase transitions. 
The regular, three-dimensional structure makes 
them also very suitable for a variety of photonic applications~\cite{SCD2016,DMO2005,HSM2012,HSC2013,KBP1984,CVH1984,N1998,ZFR2012,FZ2010,FZ2011,FYY2009,Oton20}. 
BPs are differentiated into the three sub-phases BPI, BPII and BPIII, 
depending on the structure and symmetry of their disclination network. 
BPI and BPII are characterized by a body-centered cubic or simple cubic lattice, respectively~\cite{MSA1981,TYK2015}. 
They are easily recognizable under a polarized microscope through their typical platelet texture,
which is made up of polygonal crystallites that are separated by grain boundaries.
In BPI, the disclination lines are well separated and do not intersect, whereas in BPII 
they form two interpenetrating lattices that are connected at nodal points and give rise to a tetrahedral network structure.
It is generally assumed that these topological differences between BPI and BPII 
are ultimately responsible for their different response under external strain.
BPIII, on the contrary, appears under the microscope as a blue foggy phase 
that does not exhibit the above mentioned platelet texture~\cite{HSC2011,OEP2011,GKH2016,FSO2018,Iwa2010}.
The symmetry of BPIII is the same as that of the isotropic phase, which leads to the general conclusion that
disclination lines in BPIII form random network without long-range periodicity. 
%The defects destabilise the local order structure, a feature that manifests itself in the strikingly narrow temperature range over which a pure BPIII sample is stable.

BPs exhibit shear-thinning behavior, just as other liquid crystalline phases, e.g. the smectic phase~\cite{FIKL2010}. 
Due to experimental intricacies caused by the rather limited range of temperatures over which BPs are thermodynamically stable, most of the previously conducted studies have altered the temperature (more or less rapidly) at fixed frequencies.
This allows us primarily to obtain information on the linear viscoelastic behavior during a phase transition,
but does not shed light onto the relaxation behavior and shear-induced effects, two fundamental aspects that are still not very well understood.

Most of what is known about the linear and nonlinear viscoelastic behavior of blue phases as well as their dynamic behavior under steady shear flow has been obtained numerically.
Computer simulations reveal a very rich rheological behavior, featuring periodic breakup and reformation of the disclination network 
and its rearrangement under shear~\cite{DMO2005,HSM2012,HSC2013}. 
The disclination lines can elastically withstand the flow. 
At low shear rates the apparent viscosity of the BPs is thus significantly higher than that of the isotropic or cholesteric phase. 
With increasing shear rate the disruption and reconnection of the disclination network entails  
 a periodic oscillatory shear stress on microscopic length scales. 
Further increase of the shear rate induces a realignment of disclination lines, which leads first to 
a flow-aligned Grandjean structure, and then upon further increase to a flow-aligned nematic state at the highest shear rates.

Early experimental studies reported viscoelastic properties in BPs~\cite{KBP1984,CVH1984,N1998}.  
The detailed rheological behavior of BPI and BPII was published very recently by Sahoo {\it et al.}~\cite{SCD2016}. 
They found a viscoelastic, solid-like behavior of BPI and BPII and identified various flow regimes which 
qualitatively agree with numerical predictions by Henrich {\it et al}.~\cite{HSM2012,HSC2013}.
Interestingly, they find damped oscillations in stress relaxation experiments and the formation 
of a shear flow-induced Grandjean-Cano \cite{OswaldPieranski, SCD2016} line.
All of these findings are thought to be connected to the realignment, disruption and reconnection of the disclination network in shear flow. 

In external electric fields, new BPs with different symmetry have been discovered~\cite{PS1986,Kit1990,PC1987,HKS1985,HS1988,AM2008,HMS2011,Manda20}. 
The electric fields break the cubic symmetry and induce a distortion of the lattice of DTCs. This leads to
lower-symmetry phases with orthorhombic, tetragonal, or hexagonal structures. 
At larger field amplitudes, the regular lattice breaks up and shows a transition to the cholesteric or nematic phase. 
Such non-equilibrium phase transitions in BPs are also expected in shear flow through a rearrangement 
of the disclination lines as predicted by Henrich {\it et al.}~\cite{HSM2012,HSC2013}.
All these simulation and experimental studies suggest that shear-induced, non-equilibrium structures emerge due to the rearrangement of the disclination lines.
A primary aim of the present work is to characterize linear viscoelasticity of BPI before and after shearing and to provide further evidence for shear-induced structures.

This article is organized as follows: Section \ref{methods} describes the materials and experimental methods. 
In Section \ref{results} we show rheological data and microscopy images of BPI under shear.
We concentrate especially on the effect that pre-shearing has on the viscoelasticity of BPI.
One of our findings is that BPI shows plateau moduli at high and low frequency domains, which can 
be related to the thickness of the double-twist cylinder (DTC) and the unit cell size.
The modification of the lattice structure due to pre-shear results in shear-enhanced elasticity in BPI.
We discuss the physical origin of the elasticity of BPI and finally summarize our findings in Section \ref{conclusions}.

%%%%%%%%%%%%%%%%%%%%%
\section{Experimental}\label{methods}
%%%%%%%%%%%%%%%%%%%%%

\subsection{Sample preparation}

We used a mixture of four compounds to create the blue phase I (BPI)~\cite{RSY2015}.
As a host liquid crystal, E8, PE-5CNF (4-Cyano-3-fluorophenyl 4-pentyl benzoate), and
CPP-3FF (4-(trans-4-$n$-propyl cyclohexyl)-3',4'-difluoro-1,1'-biphenyl) were mixed 
at a composition ratio of 4 : 3 : 3. 
A chiral dopant, NYC-22133L, was added at a concentration of 15 wt\%.
These compounds were obtained from LCC corporation, Japan and used as received.  
The mixture was stirred for a day at the isotropic temperature. 
It is known that this mixture forms a cubic BPI~\cite{RSY2015}.

\subsection{Methods}

The viscoelastic measurements were performed using an ARES-G2 strain-controlled rheometer from
TA Instrument Co., Ltd. with a cone - plate geometry (diameter = $25$ mm, cone angle = $0.04$ rad). 
The sample solutions were loaded on the plate at a temperature $T$ = 35 $^\circ$C corresponding to the isotropic phase.
Then the temperature was set to the measurement conditions by cooling the system.
All dynamic viscoelastic measurements were performed in the linear regime, which was confirmed 
through preliminary experiments at every temperature.
The accuracy of the temperature measurement during the rheological experiments was $\pm$ 0.02$^{\circ}$C, which ensures reliable results very close to phase boundaries.

The microscopy images in the quiescent state were taken using an Olympus BX51 
microscope with a 10x objective in cross-polarized mode.
The temperature was controlled by using a Linkam hot stage 10021. 
Sample image was observed as a thin-film sandwiched between a slide glass and cover slip.

The observations under shear were performed using a stress-controlled rheometer, 
MCR-301, Anton Paar, equipped with an Olympus inverted microscope (IX-83).
Parallel plate shear cells made of glass with diameter = 30 mm were used for the observations.
The sample thickness was kept at 50 $\mu$m. Note that this is a minimum thickness for applying shear flow. 
For thinner samples it is difficult to ensure that the glass plates are parallel.
The microscopy images under shear were obtained using a CMOS camera, HAS-L1, Ditect Ltd Co. 
Because of the different light sources in both microscope observations, the typical platelet texture of the BP 
shows slightly different colors.
For the rheological and microscopic measurements no surface treatment was performed.
In the cholesteric phase, we observed oily streaks instead of fingerprint textures (data not shown). 
The anchoring conditions on the surfaces were planar degenerate throughout.

%%%%%%%%%%%%%%%%%
\section{Results and Discussion}\label{results}
%%%%%%%%%%%%%%%%%

Figure~\ref{fig1} shows the temperature dependence of the viscosity $\eta$ at a fixed shear rate 
of $\dot\gamma$ = 1 s$^{-1}$ upon cooling as well as heating.  
The temperature was scanned in the range between 25.0 and 35.0 $^\circ$C with a scanning rate 
of $\dot T$ = 0.1 $^\circ$C/min.
The temperature dependence of the viscosity during the cooling process 
was more gradual and could be measured stably.
The viscosity increases rapidly during the cooling process at $T$ = 32.0 $^\circ$C, exhibits a shoulder at around
$T$ = 31.0 $^\circ$C and falls off at 29.0 $^\circ$C.
However, the behavior during heating was not quite as continuous as
the viscosity increased steeply at $T$ = 29.0 $^\circ$C, followed
by a gradual and slightly unsteady decrease in the temperature range between $T$ = 29.0 and 32.0 $^\circ$C.
Furthermore, a significant thermal hysteresis was observed, but only in the cholesteric phase (N$^\ast$)
and below $T$ = 29.0 $^\circ$C. This was also accompanied by different visual textures
under the polarized optical microscope, which were most noticeable in the quiescent state and became
slightly more obscured at low but finite shear rates.
The observed temperature dependence of the viscosity is similar to that seen in some other BP systems~\cite{N1998,SCD2016,FSO2018}.
It has been previously reported that the same mixture 
forms a BPI between temperatures of $T$ = 30.4 and 33.4 $^\circ$C~\cite{RSY2015}.
We therefore conclude that the observed increase of the viscosity is due to the formation of the BPI 
and attribute the pronounced hysteresis in the viscosity of the N$^\ast$ phase during the cooling process 
(viz. Fig.~\ref{fig2}) to residual effects of the BP structure.

\subsection{Identification of BP}

\begin{figure}[htpb]
 \begin{center}
	\includegraphics[width = 0.5\textwidth]{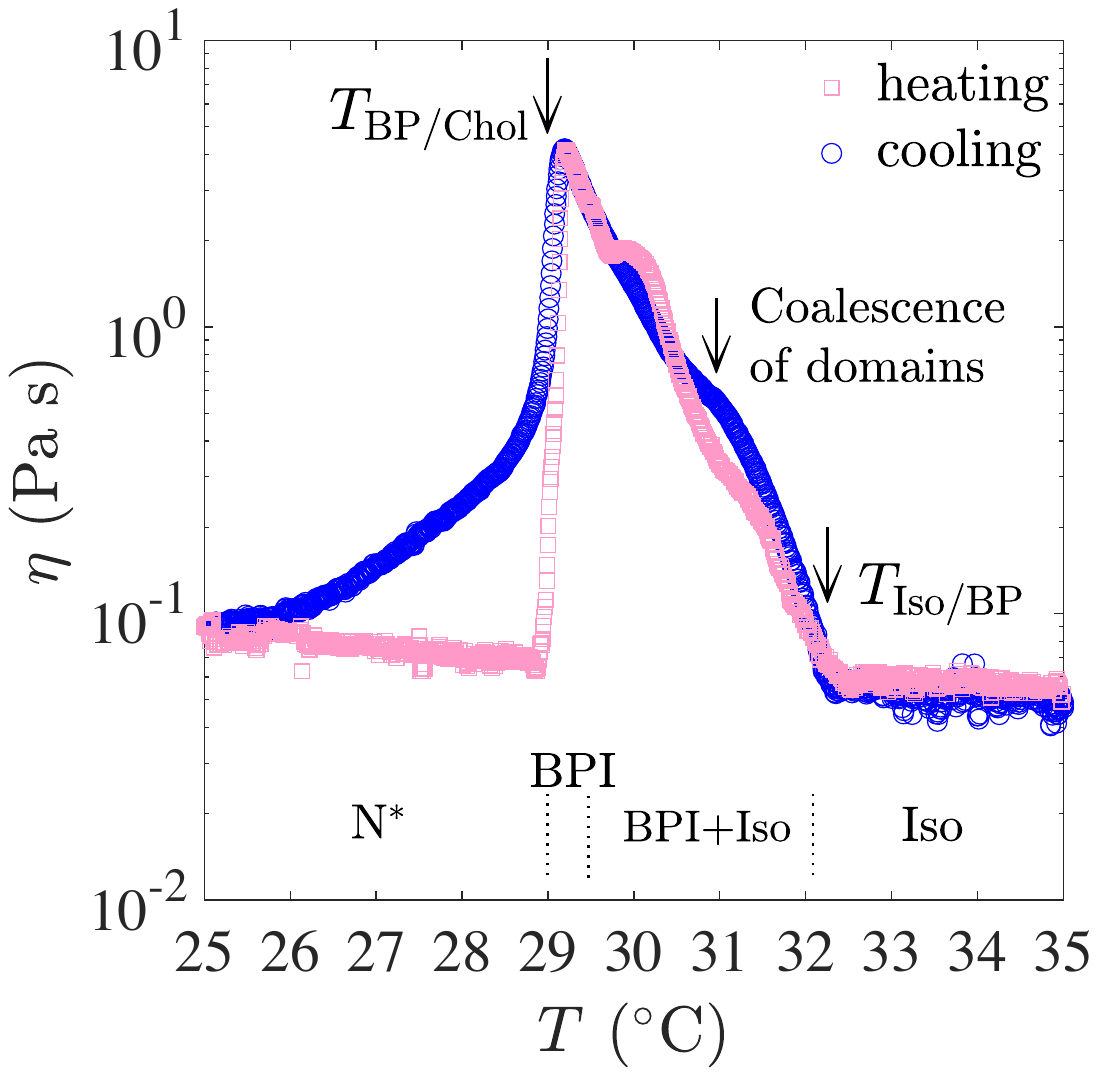}
 	\caption{
	Temperature dependence of the viscosity at $\dot\gamma$ = 1 s$^{-1}$.
	The different symbols correspond to measurements during the cooling and heating processes.
	The temperature was changed at a constant rate of $\dot T$ = 0.1 $^{\circ}$C / min.
	}
	\label{fig1}
 \end{center}
\end{figure}

\begin{figure*}[htbp]
 \begin{center}
	\includegraphics[width = \textwidth]{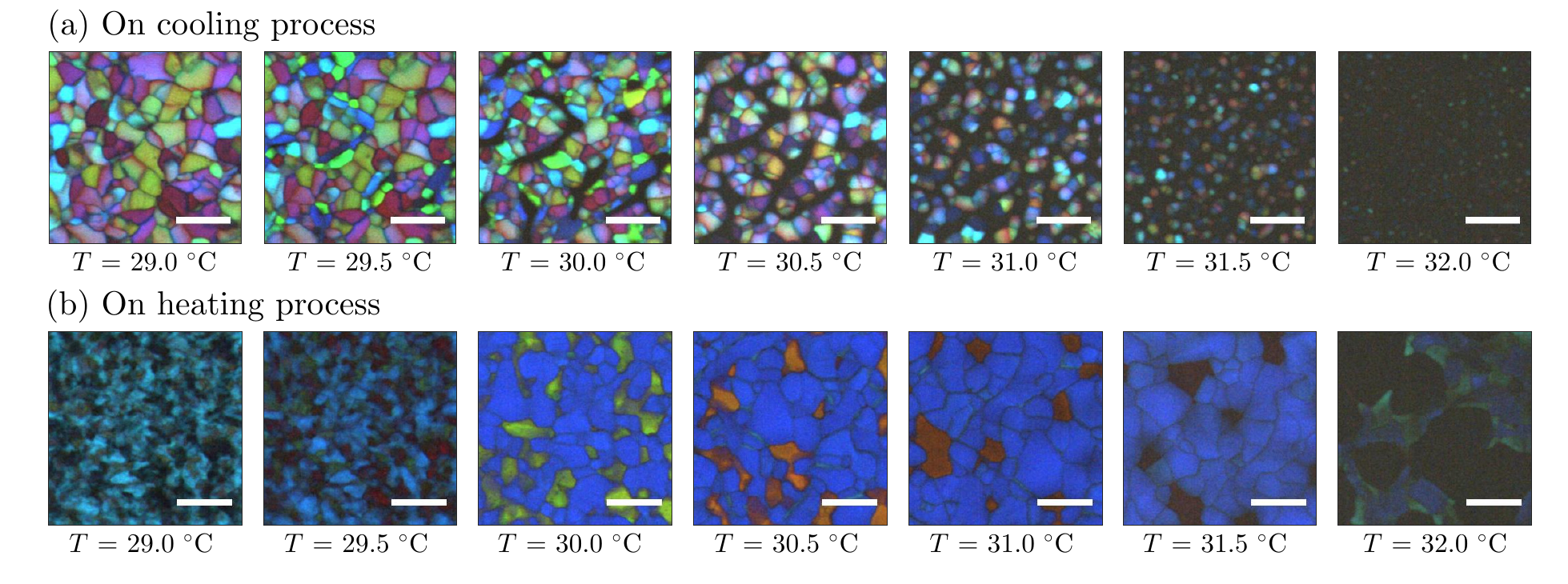}
 	\caption{
	Polarized microscope images during cooling (a) and heating (b) in the quiescent state.
	The scale bar corresponds to 10 $\mu$m.
	In each process the temperature was changed at a constant rate of 0.1 $^{\circ}$C / min.
	In the microscope observations the sample was sandwiched between a slide glass and a cover slip as a thin-film.
	}
	\label{fig2}
 \end{center}
\end{figure*}

Representative polarized optical microscopy images of the quiescent state during heating and cooling are shown in Fig.~\ref{fig2}.
During cooling, the system started in the isotropic phase at $T$ = 33.0 $^\circ$C. 
The temperature was then decreased at a rate of $\dot T$ = 0.1 $^\circ$C/min.
At $T$ = 32.0 $^\circ$C ordered structures are nucleated and grown from the dark, isotropic phase.
The size of the ordered regions increases upon further cooling and a different birefringence color pattern emerges, 
depending on the orientation of the liquid crystals.
At $T$ = 31.0 $^\circ$C spherical coalescing domains of BPI are clearly visible.
These ordered domains grow upon cooling and fill the entire sample area at $T$ = 29.5 $^\circ$C.
This is a typical feature of BPI and BPII.
During the heating process textures different from those described above are observed.
The temperature was first set to $T$ = 28.0 $^\circ$C, corresponding to the N$^\ast$ phase.
The temperature was then raised at a rate of $\dot T$ = 0.1 $^\circ$C/min.
At $T$ = 29.0 $^\circ$C, small platelet-like textures can be seen.
The size of the textures increases with temperature, and instead of a multi-colored,
polycrystalline structure a dichromic texture appears.
Pre-existing defects in the N$^\ast$ phase significantly affected this texture during heating. 
Hence, the two textures of BPI, either grown from the isotropic phase upon cooling or formed from the N$^\ast$ phase
upon heating, are obviously quite different. 
The hysteresis and unsteadiness that we observe in the viscosity measurements and
which we address more specifically below can thus be attributed directly to the distinct textures of the 
blue phase during heating and cooling.

Following Sahoo {\it et al.}, Kleiman {\it et al.} and Negita ~\cite{KBP1984,N1998,SCD2016} 
and considering the characteristic platelet texture of the BP, we attribute the sharp increase in viscosity 
between 29.0 $^\circ$C and 32.0 $^\circ$C to the formation of BPI.
On the basis of Figs.~\ref{fig1} and \ref{fig2} the phase transition temperatures in this system 
can be determined as $T_{\rm Iso/BPI}$ = 32.0 $^\circ$C for the Isotropic / BPI transition,
and $T_{\rm BPI/Chol}$ = 29.0 $^\circ$C for the BPI / N$^\ast$ phase transition, respectively.
During the cooling process, most of the BPI region is coexisting with the isotropic phase.
With regard to the temperature dependence of the viscosity, the shoulder around $T$ = 31.0 $^\circ$C
in Fig. ~\ref{fig1} can be related to the coalescing BPI domains. 
The reported phase transition temperatures between the different liquid crystalline 
phases in calorimetric measurements agree well 
with those observed in our viscoelastic measurements~\cite{KBP1984}.

\subsection{Thermal hysteresis in the shear moduli}

\begin{figure}[htpb]
 \begin{center}
	\includegraphics[width=0.5\textwidth]{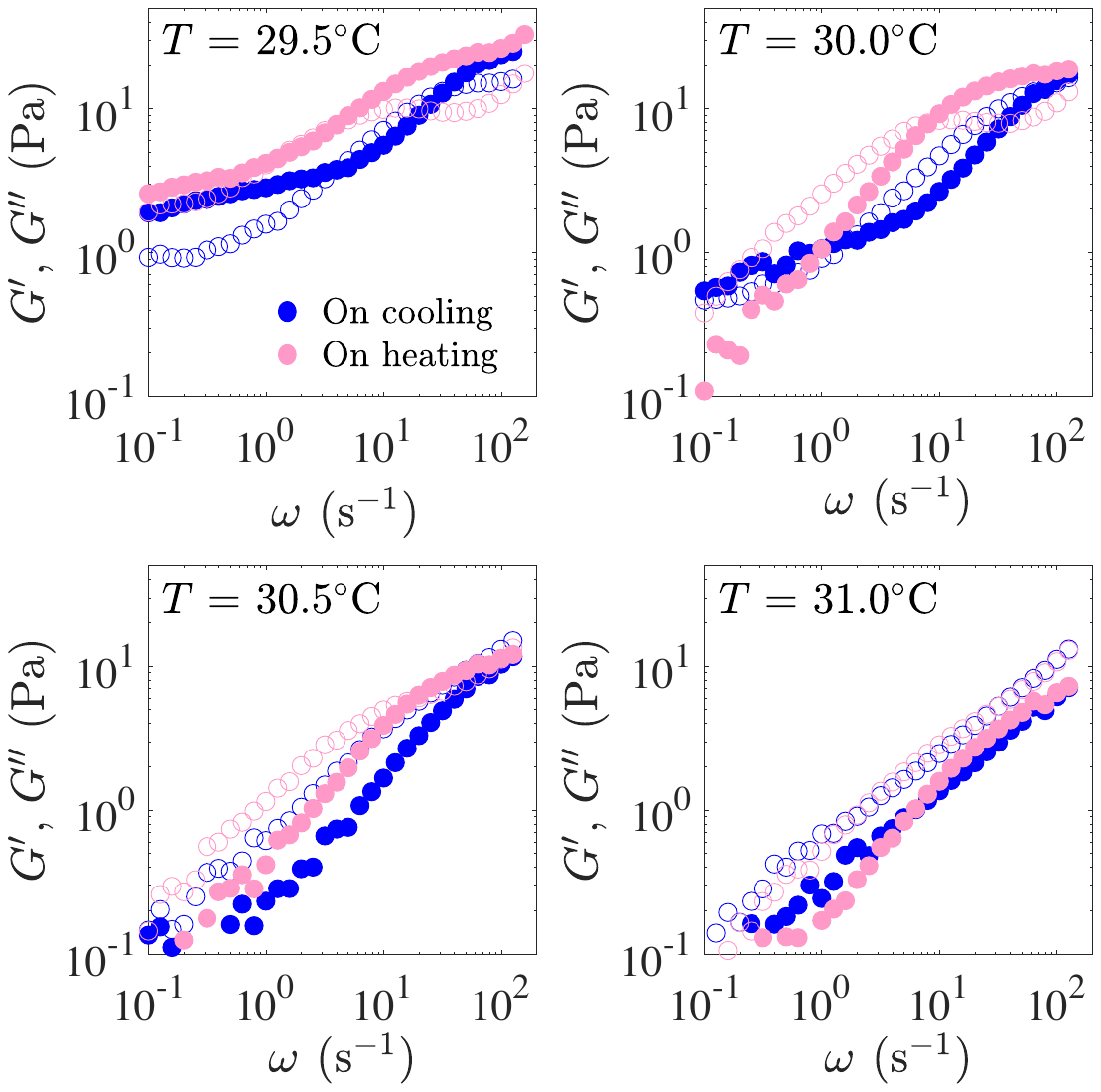}
 	\caption{
	Shear moduli $G'$ and $G''$ as a function of angular frequency $\omega$ at different temperatures.
	The shear moduli measured during the heating and cooling process are compared in the same figure.
	Filled and open symbols indicate the storage and loss modulus, respectively.
	}
	\label{fig3}
 \end{center}
\end{figure}

Our BPI system exhibits thermal hysteresis as shown in Fig.~\ref{fig2}.
In order to confirm this finding from a rheological point of view, the dynamic shear moduli 
were measured during cooling and heating. 
The temperature was changed stepwise at a rate of $\Delta T$ = 0.1 $^{\circ}$C and
was kept constant for 600 s so that the system could equilibrate. 
Then the dynamic shear moduli were measured as a function of angular frequency, 
followed by the next temperature increment or decrement.
This procedure was repeated for temperatures between $T$ = 32.0 and 28.5 $^\circ$C.
The shear moduli at representative temperatures are shown in Fig.~\ref{fig3}.
At $T$ = 31.0 $^\circ$C, corresponding to the BPI/Iso biphase region, storage and loss modulus $G'$ and $G''$ 
were measured upon cooling and heating. Interestingly, both thermal cycles nearly coincide.
$G''$ dominates over the entire frequency domain, which characterizes the BPI/Iso biphase system as a viscous fluid.
The hysteresis in the dynamic shear moduli appears gradually as the temperature is decreased.                                  
At $T$ = 30.5 and 30.0 $^\circ$C, the viscoelasticity during the heating cycle shows a plateau modulus in $G'$ 
at high frequencies. However, the viscoelastic profile is still dominated by $G''$, 
suggesting fluid-like behavior is more pronounced than elasticity.
On the other hand, during the cooling process the shear moduli 
show the viscoelastic, solid-like behavior at $T$ = 30.0 $^\circ$C, visible in a secondary plateau in the modulus 
at low frequencies. This has been observed previously in other BPI and BPII systems~\cite{SCD2016}.
At $T$ = 29.5 $^\circ$C, the second plateau modulus at low frequency forms during both cooling and heating 
with only a slight difference in the frequency dependence. 
Hence, we attribute the viscoelastic, solid-like nature of this phase to the formation of a homogeneous BPI.
The shear moduli of the homogeneous BPI appear to be different from those reported by Sahoo et al.~\cite{SCD2016}. We would like to mention that the apparently differences can be attributed to the temperature dependence of the relaxation time. As shown in Figure~\ref{fig1}, cooling the system within the BPI phase region increases its viscosity due to the increase in relaxation time. A BPI phase with long relaxation time will exhibit a dynamic viscoelastic behavior like that presented by Sahoo et al.

The thermal hysteresis behavior has been reported before during phase transitions of BPs~\cite{CM1978},
which are also known to show electro-optical hysteresis ~\cite{CGX2010,CLW2012}. In all instances 
the observed hysteresis has been directly related to the growth and orientation of blue phase crystallites,
suggesting the different visual textures during the heating and cooling process are responsible for the dynamical hysteresis
and the differences in rheological behavior that we observe.

\subsection{Shear hysteresis in the shear moduli}

In the following, we present only the experimental data that was obtained for the cooling cycle.
As previously demonstrated by numerical studies, the realigning of the network of disclination lines in shear flow 
can have a determining influence the rheological properties of BPI~\cite{DMO2005,HSM2012,HSC2013}. 
Here, we compare the effect of pre-shear on the BPI/Iso biphase and the homogeneous BPI phase.

\subsubsection{BPI/Iso biphase region}

\begin{figure}[htpb]
 \begin{center}
	\includegraphics[width=0.5\textwidth]{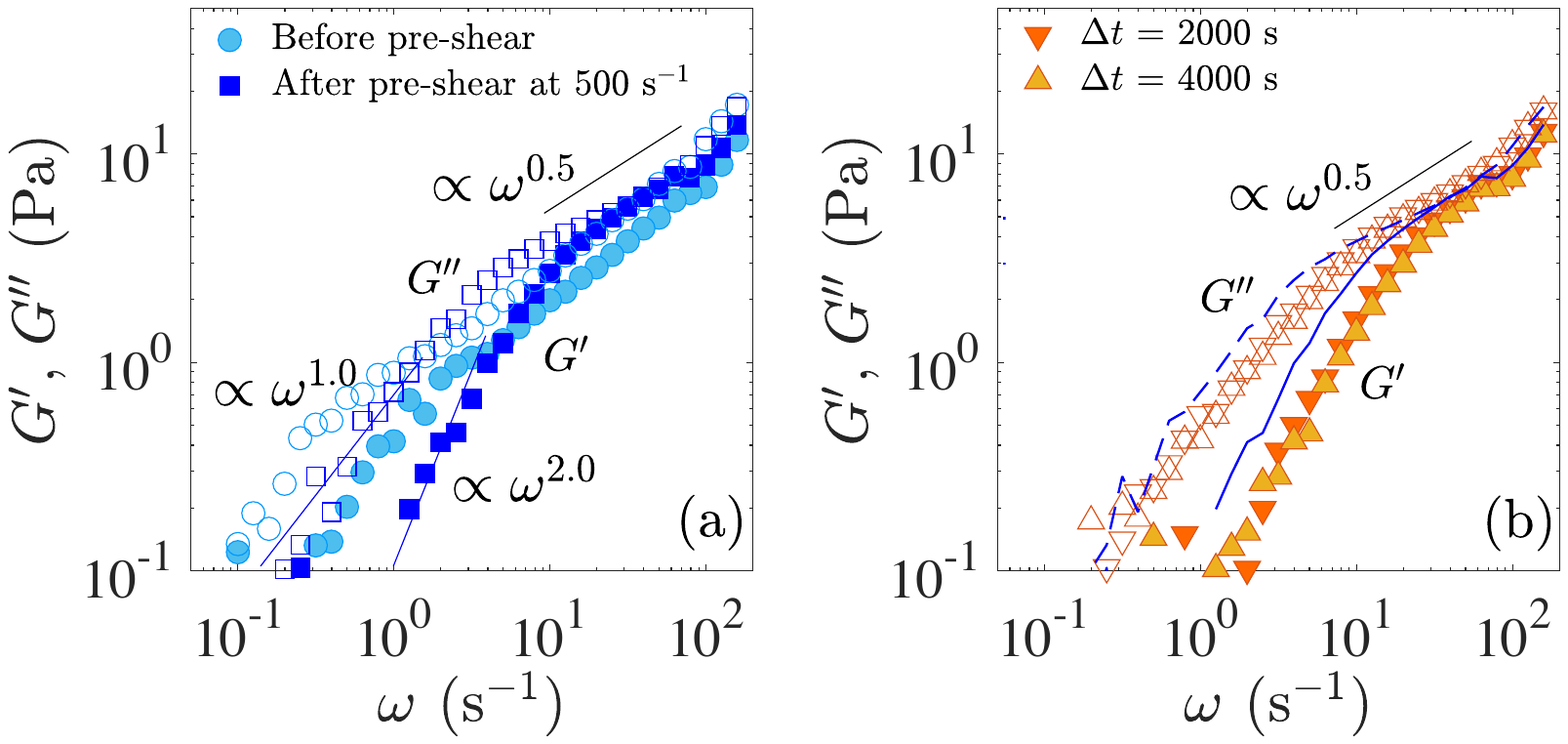}
 	\caption{
	Shear moduli of BPI/Iso biphase at $T$ = 31.0 $^{\circ}$C as a function of angular frequency $\omega$: 
	Shear moduli measured before and after applying pre-shear rate of $\dot\gamma_{\rm pre}$ = 500 s$^{-1}$ are
	compared in panel (a).
	Solid lines indicate the slope of the shear moduli $\omega^{0.5}$,  $\omega^{1.0}$ and  $\omega^{2.0}$ 
	for the reference, respectively. 
	Shear moduli measured at different elapsed times $\Delta t$ = 2000 s and 4000 s 
	after pre-shear are also shown in panel (b). 
	Filled and open symbols are storage and loss shear modulus, respectively.
	Solid and dashed lines correspond to the shear moduli after pre-shearing at $\dot\gamma_{\rm pre}$ = 500 s$^{-1}$.
	}
	\label{fig4}
 \end{center}
\end{figure}

Fig.~\ref{fig4}(a) shows the shear moduli in the BPI/Iso biphase region at 
$T$ = 31.0 $^\circ$C before and after the application of pre-shear with a shear rate of $\dot\gamma_{\rm pre}$ = 500 s$^{-1}$.
Without pre-shear the shear moduli $G^\ast$ obeys a power law behavior like 
$G^\ast \sim \omega^{1/2}$ in the high frequency domain.
We presume this is caused by viscous dissipation between BPI domains that are formed in the isotropic phase as 
it occurs also for concentrated emulsions~\cite{LRM1996}.
The absence of a plateau in the storage modulus $G'$ and the dominance of the loss modulus $G''$ 
gives evidence that the BPI/Iso biphase behaves like a viscous fluid. 
After applying a pre-shear at $\dot\gamma_{\rm pre}$ = 500 s$^{-1}$ the moduli  
are slightly enhanced in the high frequency domain whilst they still maintain 
the above mentioned power law behavior. 
The storage and loss moduli in the terminal region are reduced and feature a single relaxation behavior
like $G' \sim \omega^{2}$ and $G'' \sim \omega^{1}$, respectively. 
The modified slope in frequency dependence of the dynamic moduli after pre-shearing 
indicates that the relaxation dynamics of the BPI is either affected
by a realignment of the disclination lines or by the emergence of a polycrystalline texture.
In Fig.~\ref{fig4}(b) the shear moduli after a specified elapsed time interval are shown.
After waiting times of $\Delta t$ = 2000 and 4000 s the single relaxation-like behavior remains unchanged, but
the shear moduli are smaller and do not recover their initial behavior. 
This is consistent with the system adopting a new (meta)stable state.
A Maxwellian single relaxation behavior that we observe after pre-shearing suggests 
that the platelet size is homogenized during the shearing.

\subsubsection{Homogeneous BPI}

\begin{figure}[htpb]
 \begin{center}
	\includegraphics[width=0.5\textwidth]{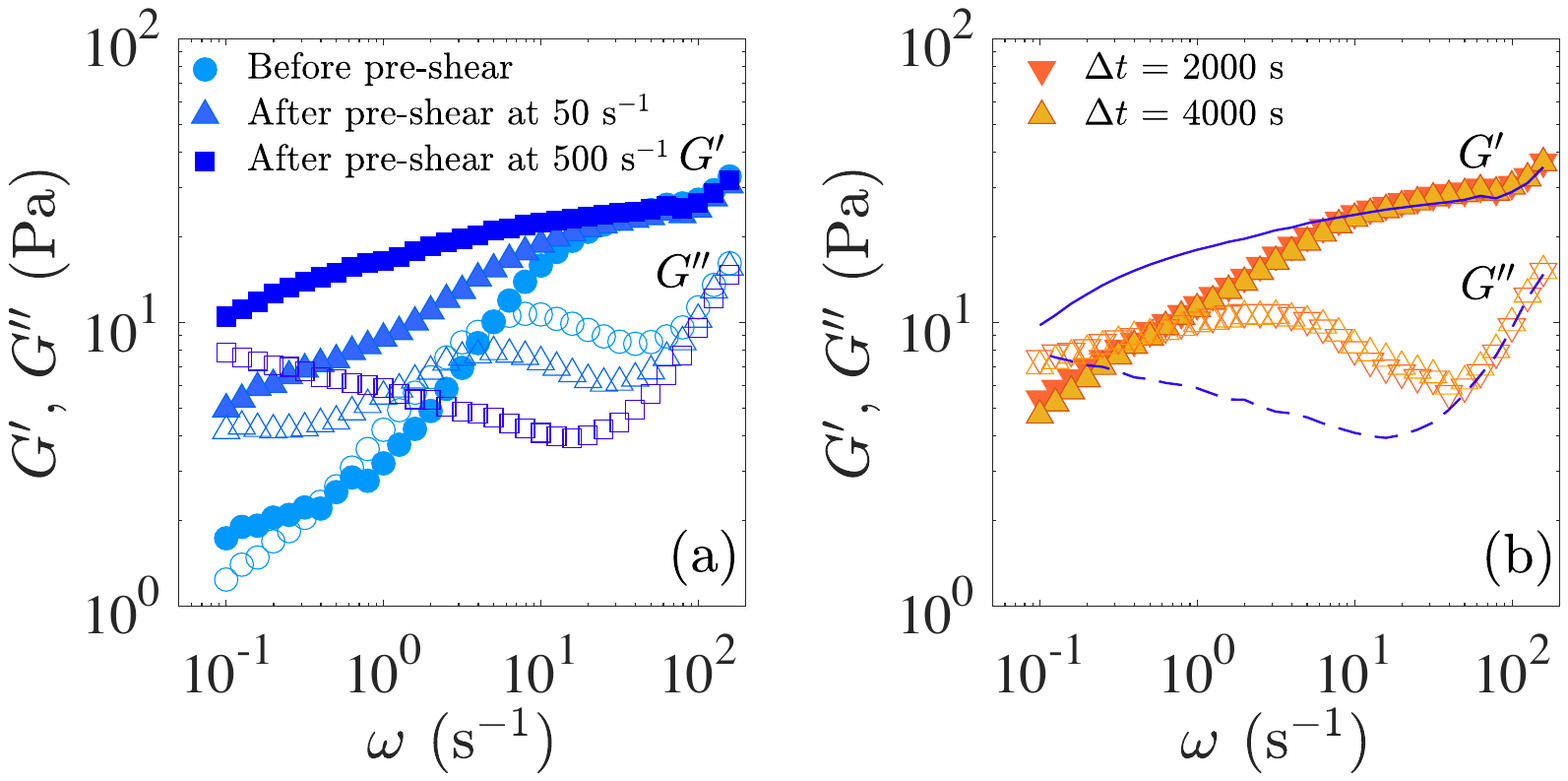}
 	\caption{
	Shear moduli of homogeneous BPI at $T$ = 29.5 $^{\circ}$C as a function of $\omega$ measured 
	before and after applying pre-shear rate of $\dot\gamma$ = 50 and 500 s$^{-1}$:
	(a) Representative results of the pre-shear effect on the shear moduli are summarized.
	(b) Shear moduli measured at $\Delta t$ = 2000 s and 4000 s after pre-shearing at 
	$\dot\gamma_{\rm pre}$ = 500 s$^{-1}$ are shown. 
	Filled and open symbols are storage and loss shear modulus, respectively.
	Solid and curved lines correspond to the shear moduli after pre-shear at $\dot\gamma_{\rm pre}$ = 500 s$^{-1}$.
	}
	\label{fig5}
 \end{center}
\end{figure}

The homogeneous BPI phase shows pronounced viscoelasticity (viz. Fig.~\ref{fig5})
contrary to the BPI/Iso biphase system.
We found a significant effect of pre-shearing on the dynamic moduli in the homogeneous BPI phase. 
Pre-shear flow induces an enhancement of the elastic modulus and slows down the relaxation. 
The enhanced elasticity does not relax towards that of the original state, 
but remains high even after the cessation of shear. 
Instead of the typical platelet texture of BPI, we observe a homogeneous texture in the state with enhanced elasticity. 
In the following, we show details of the pre-shear effect on the shear moduli.

Before applying the pre-shear, in Fig.~\ref{fig5}(a), the storage modulus $G'$ has a plateau modulus at high frequencies.
One can also see a second plateau in $G'$ at low frequencies, supporting 
the existence of a numerically predicted yield stress~\cite{HSC2013}.
The loss modulus $G''$ exhibits a minimum around $\omega$ = 40 s$^{-1}$.
For intermediate frequencies at about $\sim \omega$ = 10 s$^{-1}$ a maximum in $G''$ appears, 
which is a clear sign of a structural relaxation.
The power law behavior of $G^\ast \sim \omega^{1/2}$ is not observed in the homogeneous BPI.
The fact that the modulus does not vanish at $\omega$ = 0 suggests BPI behaves 
thermodynamically like a solid phase with a finite yield stress.

Applying pre-shear changes the viscoelasticity of BPI drastically.
While the plateau modulus at high frequencies remains unchanged, the minimum and maximum in $G''$ shift to lower frequencies. 
At low frequencies the pre-shearing has an even more remarkable influence
as the shear moduli are significantly enhanced with increasing pre-shear rate. 
Particularly after applying a rate of 500 s$^{-1}$ the storage modulus $G'$ dominates 
the entire frequency domain.
With increasing pre-shear rate the homogeneous BPI phase becomes more elastic and features longer relaxation times.
As the consequence of the significant slowing down of the structural relaxation, the maximum in $G''$ shifts gradually out of the window that we can access experimentally towards lower frequencies.
This monotonic behavior of $G''$ is a signature of the structural relaxation of BPI.

Fig.~\ref{fig5}(b) shows the shear moduli measured for waiting times of $\Delta t$ = 2000 s and 4000 s after 
the application of the pre-shearing, which gives rise to further change in the shear moduli.
The initial state is not fully recovered and the shear-enhanced elasticity that we report here
is therefore irreversible. 
The persistent nature of the dynamic shear moduli indicates that a shear-induced phase has formed in BPI.

During all rheological measurements reported here, the temperature was stable and accurately measured as mentioned in the experimental section, and the stability of the homogeneous BPI phase was retained.

\begin{figure*}[htpb]
 \begin{center}
	\includegraphics[width=\textwidth]{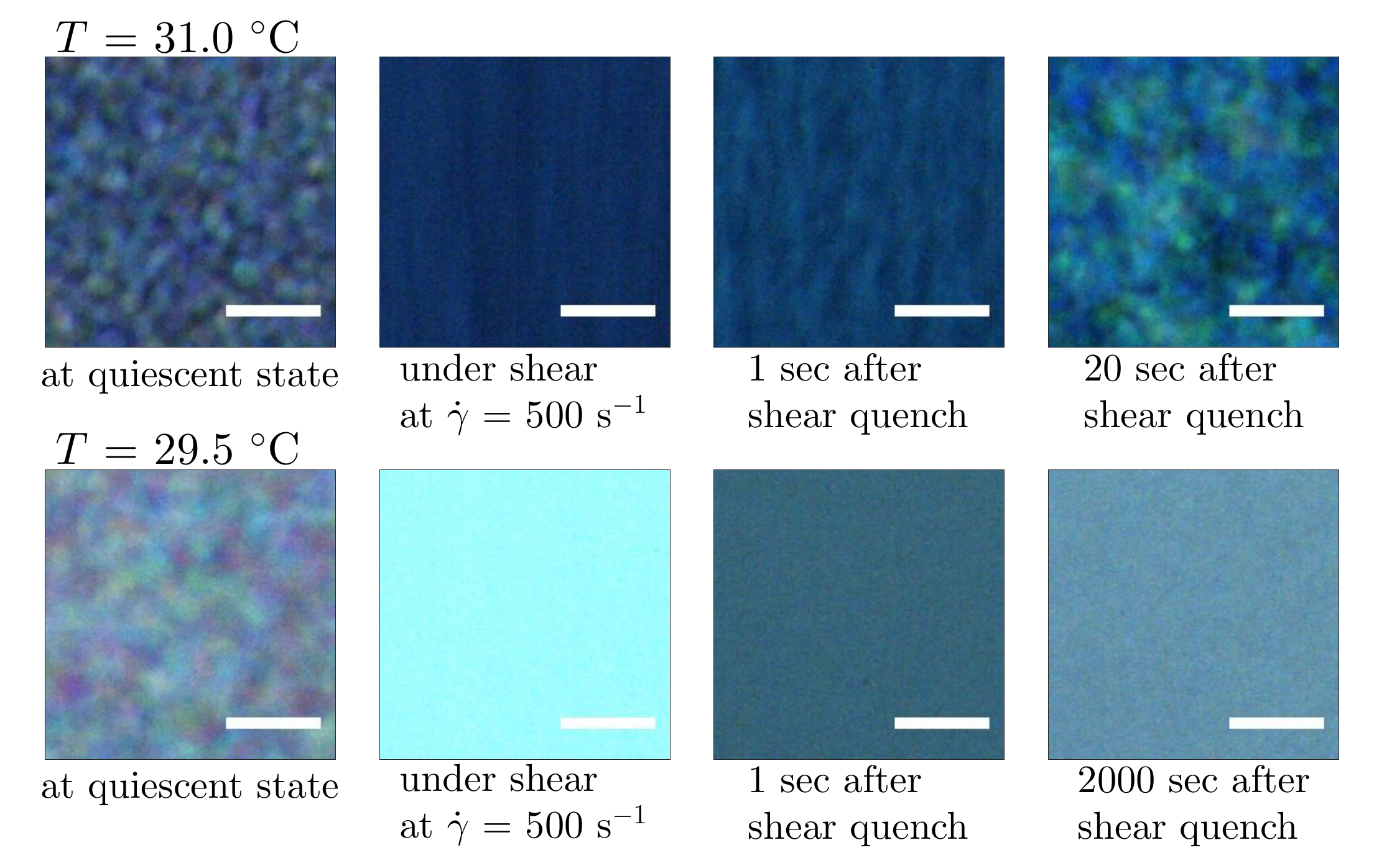}
 	\caption{
	Polarized microscopy images of BPI and BPI/Iso biphase taken after a quench from 
	shear rates of $\dot\gamma$ = 500 s$^{-1}$ to 0 s$^{-1}$.
	The images in the upper row show results at $T$ = 31.0 $^\circ$C, whereas 
	those in the lower row give results at $T$ = 29.5 $^\circ$C.	
	The scale bar corresponds to 10 $\mu$m and the flow is along the vertical direction.
	}
	\label{fig6}
 \end{center}
\end{figure*}

The textures under shear flow at different temperatures are shown in Fig.~\ref{fig6}.
Due to the larger thickness of the sample that is required for applying homogeneous shear flow 
the fraction of scattered light is unavoidably larger. This leads to a slightly more 
blurred appearance of the images. Their polycrystalline structure, however, is still clearly recognizable
and demonstrates that we are still operating in the BPI region.
Applying a pre-shear rate of $\dot\gamma_{\rm pre}$ = 500 s$^{-1}$ at $T$ = 31.0 $^{\circ}$C 
breaks up the platelet texture in the BPI/Iso biphase, which is then immediately recovered upon cessation of the shear flow.
%Immediate recovery of the platelet texture in BPI/Iso biphase indicates that the orientation of the disclination lines 
%rearranged by shear flow are easily disturbed by a nucleation and growth of the platelet texture. 
At a lower temperature of $T$ = 29.5 $^{\circ}$C we observe a remarkable irreversibility of the BPI texture 
as with the same pre-shear applied the texture is completely broken and does not recover.
In the thin sample, one can see that the platelet texture is dissolved when the force is applied on the BPI \rev{\cite{SMMovie}}.

The polarized optical microscopy images of BPI at $\dot\gamma$ = 500 s$^{-1}$ are generally brighter than those of the BPI/Iso biphase. 
In the BPI/Iso biphase, the dark polarized image under shear flow indicates a collapse of ordered domains. 
On the contrary, the bright image in the homogeneous BPI suggests a reorientation of the director field and presumably also of the DTC network,
indicating a structural transition induced by the shearing. 
We would like to emphasize that because the value of the first plateau modulus in $G'$ at high frequencies, which is attributed to the DTCs, is virtually the same before and after pre-shearing as evident from Fig. ~\ref{fig5} (b), the DTC structure must remain largely intact. It is, however, not possible to draw any strict conclusions about the exact orientation of the DTCs from these findings. We also believe that the DTCs are certainly not perfectly oriented.

The reduction in birefringence after the cessation of shear suggests that the flow-oriented DTCs relax.
In the homogeneous BPI, even after cessation of the shear flow the platelet texture is not recovered. 
The fact that we do not observe any distinct textures like platelets suggests that 
the nucleation is prohibited and that the BPI is arrested in a shear-induced state (see also comment below in Section \ref{shear-enhanced-elasticity}).
Here, we would like to remark that a focal conic texture as commonly observed in N$^{\ast}$ phases is not detected. 
This gives evidence that BPI is the thermodynamically stable phase and a phase transition to the N$^{\ast}$ phase is not induced
under shear in our experimental condition.
For thin enough samples BPI shows a dark foggy phase after the shear quench \rev{\cite{SMMovie}}. 
From polarized optical microscopy it is known that BPIII has a foggy texture because of the amorphous alignment of the disclination lines.
The foggy phase after shearing BPI suggests that the ordered alignment of the disclination lines is disturbed and a network of disordered disclination lines is formed.

\subsection{Origin of the elasticity}

As previously stated the pre-shearing breaks up the platelet texture at a temperature $T$ = 29.5 $^\circ$C, 
which is also where we observe the transition to a homogeneous BPI. 
Interestingly, the system does not recover the initial state, which is true even after the shear flow has been completely 
switched off, as shown in Fig.~\ref{fig6}. 
Therefore, the shear-enhanced elasticity has to originate necessarily from the rearrangement of three-dimensional array 
of DTCs and disclination lines, and not from the re-orientation and compression of the platelet texture.
As suggested by numerical studies, the disclination lines can withstand a certain degree of deformations in 
the shear flow. This observed resistance constitutes at least one possible contribution to the overall elasticity 
of the BP ~\cite{DMO2005,HSC2013} and may well be the origin of a yield stress.
Flow curves measured by using the strain-controlled rheometer, ARES-G2, and apparent yield stress estimated from them are shown in Fig. 1S of the Supplemental Material of this paper ~\cite{SM}.
In Fig. 1S, the apparent yield stress is also compared with the second plateau modulus $G'$ at $\omega$ = 0.1 s$^{-1}$.

Evidence in support of a characteristic dynamical length scale has been given by 
Fukuda {\it et al.}~\cite{FZ2011b}. 
In their numerical study of the structural forces that are mediated by the BP ordering 
under both compression and dilation, they found that because of the periodicity of BPI 
the force shows an oscillatory behavior with minima at a dilation or compression 
length that equals half a unit cell size of a BPI structure.
Following a simple argument~\cite{SRCF}, it can be shown that the shear modulus scales like 
$G_{0}\sim k_{\rm B}T/\xi^3$, where $\xi$ has the meaning of  
a characteristic length scale, such as the lattice constant of a periodic structure.
Using the high frequency plateau modulus $G'\simeq$ 30 Pa at $T$ = 29.5 $^\circ$C 
(viz. Fig.~\ref{fig5} left image) this leads to an estimate of $\xi\simeq$ 50 nm.
Using the same reasoning and the value of the modulus $G'\simeq$ 2 Pa 
at the point where it levels off at low frequencies, this yields a value of $\xi\simeq$ 130 nm,
 which is approximately half the value that has been reported for the lattice 
constant~\cite{Crooker,TYK2015,MG1996,CH2015}.

Through transmission electron microscopy (TEM), the periodicity of the DTC lattice in BPI was
accurately determined to be 238 $\pm$ 10 nm for the mixture of reactive mesogen RMM-141C (Merck) 
and the strongly twisting chiral dopant CD-X (Merck)~\cite{TYK2015}.
The dependence of the lattice constant on the chiral dopant for different mixtures has been also studied 
by reflection spectra and Kossel diagrams, which led to slightly larger values in the range of 280 - 300 nm,
showing good agreement between different mixtures\cite{CH2015}.  
It is interesting that our value of the characteristic length scale $\xi$ in the rheological experiment  
is strikingly close to the diameter of a DTC in a BPI lattice and only about a factor two to four
smaller than the lattice constant. It should be emphasized that some sort of deviation is hardly surprising, 
given that the above value of $\xi$ was obtained by different procedures, i.e. through the dynamical response of BPI.
The shift to slightly lower apparent values can also be motivated through the fact that the array of DTCs 
is not completely rigid, but also subject to a certain degree of deformation during shearing.

This corroborates our idea that the length scale $\xi$ of half a unit cell is indeed 
the relevant rheological structural unit for the elastic response to shear forces
and that the angular frequency dependence of the storage modulus has to originate from the lattice of DTCs. 
The peak in the loss modulus $G''$, which appears at frequencies between the low- and the high-frequency plateau 
regions in the storage modulus $G'$, indicates as well a transition from a high-frequency regime where the 
dynamical response of DTCs is dominant to another one at lower frequencies that is mainly influenced by the 
length scale of the disclination lattice. This allows us to interpret the relaxation time 
$\tau_{\rm p}$, in our approach estimated through the peak frequency in $G''$, as dynamic signature of the DTCs.
The fact that the second plateau modulus corresponds approximately to the apparent yield stress suggests 
that the distortion of a lattice unit cell of BPI results in the regular 
array of DTCs and disclination lines being broken up and adopting a shear-induced conformation.

\subsection{Shear-enhanced elasticity}\label{shear-enhanced-elasticity}

\begin{figure}[htpb]
 \begin{center}
	\includegraphics[width=0.5\textwidth]{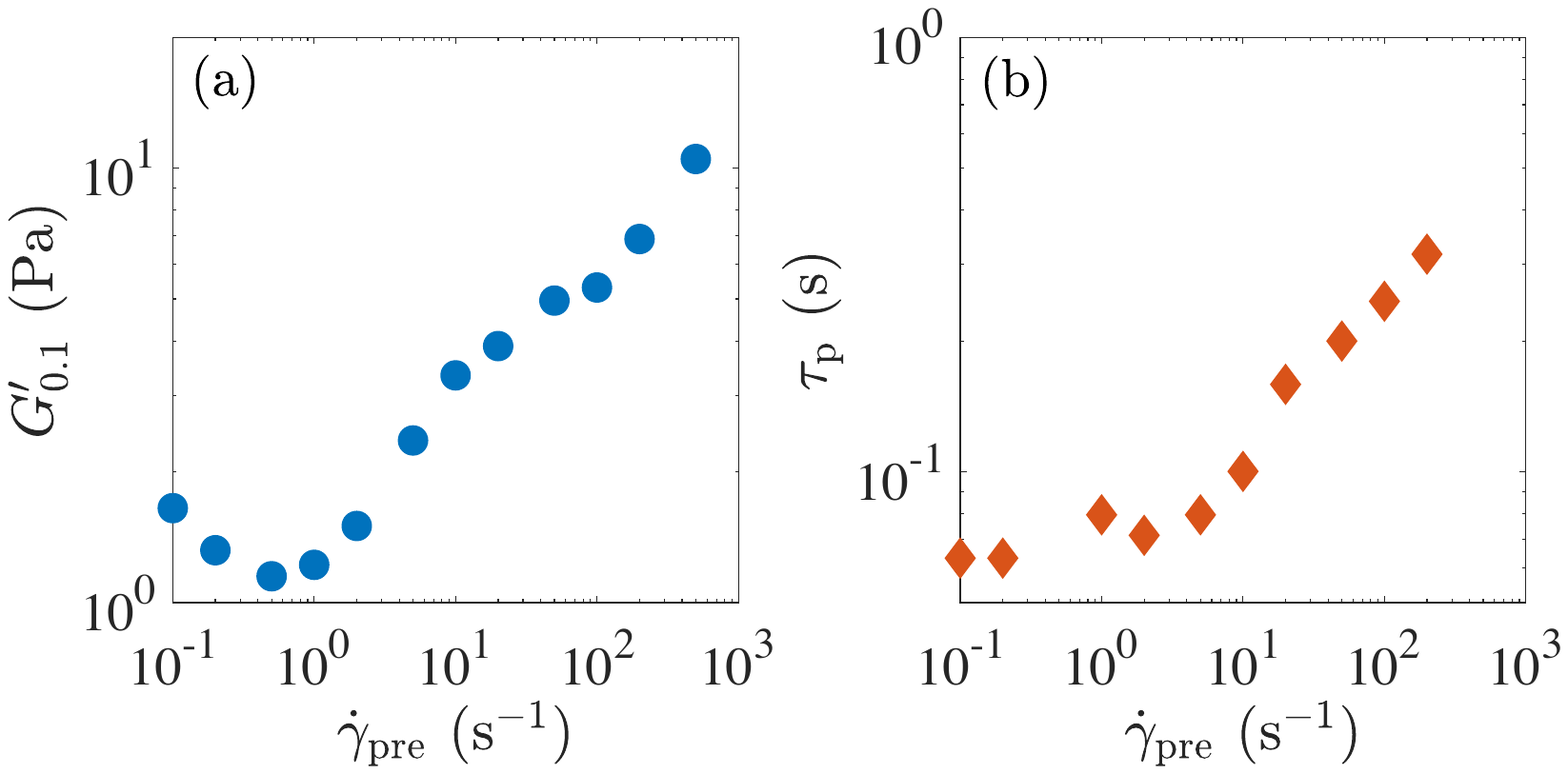}
 	\caption{
	Effect of the pre-shear treatment on the shear modulus $G'$ and relaxation time in BPI at $T$ = 29.5 $^{\circ}$C. 
	Here, the shear modulus at a representative frequency of $\omega$ = 0.1 s$^{-1}$ is shown.
	The relaxation time $\tau$ was estimated as the inverse frequency where the peak 
	in the loss shear modulus $G''$ was observed. 
	}
	\label{fig7}
 \end{center}
\end{figure}

Fig.~\ref{fig7} shows the dependence of the storage modulus $G'$ on the pre-shear rate 
$\dot\gamma_{\rm pre}$ at $\omega$ = 0.1 s$^{-1}$.
This frequency was chosen because the most pronounced effect of pre-shearing was observed
for this value. The relaxation time $\tau_{\rm p}$ is estimated as the inverse of the frequency of the maximum in the loss modulus $G''$. 
 
At low shear rates $\dot\gamma_{\rm pre}$ $<$ 1 s$^{-1}$ the storage modulus $G'_{0.1}$ decreases with $\dot\gamma_{\rm pre}$ 
and the relaxation time $\tau_{\rm p}$ remains almost constant. 
However, above a critical shear rate in the range of $\dot\gamma_{\rm pre}$ $>$ 0.5 - 1 s$^{-1}$, 
$G'_{0.1}$ and $\tau_{\rm p}$ both increase with increasing $\dot\gamma_{\rm pre}$.
Note that $G'_{0.1}$ and $\tau_{\rm p}$ increase even without exhibiting visible platelet textures, as shown in Fig.~\ref{fig6}.
The increase of $G'_{0.1}$ and $\tau_{\rm p}$ indicates that a realignment of the disclination lines 
contributes primarily to the slowing down of the relaxation time and thus to the elasticity of the investigated BPI. 
The findings also suggest that a critical shear rate needs to be applied to trigger the rearrangement process.
The distortion of the lattice unit cell induces the re-ordering of the disclination lines. 
But due to the effect that pre-shear has on the measured shear modulus (viz. Fig.~\ref{fig7}(b)), the relaxation time 
$\tau_{\rm p}$ remains almost constant at low shear rates $\dot\gamma_{\rm pre}$ $<$ 1 s$^{-1}$.
This insensitivity towards can only be rationalized if no drastic realignment of the DTC and the 
disclination network takes place.

A more significant rearrangement of the disclination lines occurs obviously above shear rates $\dot\gamma$ $>$ 1 s$^{-1}$,
suggesting that the shear-induced state remains arrested. 
Even after longer waiting times the system does not revert to its original structure, nor is there any precursor observable as change in its rheological properties (viz. \ref{fig5} (b)). As pointed out in \cite{Henrich2010}, the difference in free energy density between BPs is typically of the order of 100 to 1,000 $k_B T$ per unit cell, even for a rather small BP lattice constant of 160 nm. Such unit-cell differences represent a very reasonable estimate of the barrier to topological reconnections within evolving BP structures and are much larger than thermal energies.
However, external driving forces like shear flow are well capable of overcoming these barriers and allow the system to access states that would be otherwise inaccessible by thermal forces alone. These states are shear-induced, but metastable in the sense that they are out of equilibrium and do not represent states with the lowest free energy upon cessation of the shear flow. Owing to the large topological barriers they can be considered permanently arrested.
 
It appears that the peak of $G''$ broadens with increasing pre-shear rate (Fig.~\ref{fig5}(a)), which is usually interpreted 
as a distribution of relaxation times.
Hence, the broadening of the peak suggests there is a multitude of shear-induced states in which the disclination network can get arrested.

Dimensionless Ericksen numbers express the ratio of viscous to elastic forces and are defined 
as $Er=\eta\nu l/K$ with $\eta$ as shear viscosity, 
$\nu$ as typical flow velocity, $l$ as typical length scale and $K$ as elastic constant of the liquid crystal.
Here, we use $\xi$ = 130 nm as characteristic length scale $l$ and estimate $\nu$ as the velocity difference 
across half a unit cell, i.e. $\nu$ = $\dot\gamma\xi$.
The value of $\eta$ was taken from experiments at shear rates $\dot\gamma$ $>$ 1 s$^{-1}$ and assumed to be 
$\eta\simeq$ 3.5 Pa s. The elastic constant $K$ = 3 $\times$ 10$^{-12}$ N was matched to the value chosen 
by Meiboom {\it et al.}~\cite{MSA1981}. This leads to an Ericksen number $Er\simeq 0.02$ 
for shear rates around $\dot\gamma$ = 1 s$^{-1}$.

In numerical studies a flow regime was found at Ericksen numbers $Er$ $\simeq$ 0.02, where shearing leads 
to periodic break-up and reformation of the disclination network which eventually cause 
the flow-induced transition of the regular network into an amorphous network state~\cite{HSC2013}.
The disclination network of BPI underwent significant conformational changes even for very 
low shear rates. Interestingly, the flow-induced amorphous state of the disclination network remained arrested 
in a metastable state and forms for Ericksen numbers up to $Er \leq$ 4.
These simulations also predicted that the system adopts another flow-induced conformation at 
high shear rates in the range of the Ericksen number 5 $\leq$ $Er$ $\leq$ 16,
a so-called Grandjean texture, which consists of a cholesteric helix with the helical axis oriented 
along the direction of the flow velocity gradient. Eventually, at very large shear rates 
corresponding to $Er >$ 16, a transition from the Grandjean 
texture to a flow-aligned nematic state takes place. 

In this experimental study Ericksen numbers were considerably lower in the region of 
0.01 $\leq$ $Er$ $\leq$ 0.33 at $T$ = 29.5 $^{\circ}$C.
However, the increase of the shear modulus and relaxation times as a function of the pre-shear rate 
(viz. Fig.~\ref{fig7}) fits well into the general picture of a shear-induced break-up 
of the disclination network and an amorphous flowing state of BPI.
Our polarized microscopy images of the homogeneous BPI at high shear rate, shown in Fig.~\ref{fig6}, 
support this interpretation and are compatible with how an amorphous flowing network of defect lines
would appear under the microscope. This suggest there is a shear-induced phase in BPI, 
similar to those observed in other defect-mediated soft matter systems 
like thermotropic smectic phases and lyotropic lamellar phases.

%%%%%%%%%%%%%%%%%%%%
\section{Conclusions}\label{conclusions}
%%%%%%%%%%%%%%%%%%%%/

The shear flow of blue phases is a representative example of structural rheology of soft condensed matter, 
which is mediated and determined by the dynamics of disclination lines and defects. 
We studied the linear rheological behavior of the cubic blue phase I (BPI).
The dynamic shear modulus $G'$ of BPI features a first and second plateau at high and low frequencies.
The simple scaling argument for the plateau shear modulus $G_0\sim k_{\rm B}T/\xi^3$ suggests that 
characteristic length estimated from high frequency plateau $\xi\sim$ 50 nm is close to a typical value of diameter of DTC.
On the other hand, the length scale obtained from the second plateau at low frequency $\xi\sim$ 130 nm approximately 
corresponds to half the lattice constant in BPI. 
The structural components responsible for each plateau modulus can be related to the diameter of single DTC and 
half the size of the lattice unit cell.
We also found that the second plateau modulus almost corresponds to the apparent yield stress. 
This correspondence indicates that the length scale of half the unit lattice cell is relevant rheological structural unit 
for the elastic response. 
The origin of the elasticity in BPI can be therefore attributed to the body centered (BC) lattice structure consisting 
of an array of double twist cylinders (DTCs).

Another interesting finding in this study is that $G'$ of a homogeneous BPI is significantly enhanced by application 
of a pre-shearing protocol.
As the modulus is enhanced with pre-shearing, the relaxation time also slows down. 
The shear-enhanced elasticity is observed above a pre-shear rate of $\dot\gamma_{\rm pre}\simeq$ 1 s$^{-1}$, 
which corresponds to the Ericksen number of $Er\simeq$ 0.02.
After pre-shearing, the platelet texture melts into homogeneous phase without visible texture. 
No focal conic texture in the shear-induced phase assures that the system is in BPI but not in shear-induced N$^{\ast}$ phase. 
The shear-enhanced elasticity above the critical shear rate $\dot\gamma_{\rm pre}\simeq$ 1 s$^{-1}$ stems from 
the disclination lines that become realigned by the shear flow and then arrest in a metastable state, where the large energetic barriers associated with topological reconnections prevent the system from reaching its equilibrium state.
If the shear-induced phase has an amorphous network of the disclination lines, it will be interesting to compare the rheological
behavior of BPIII, which has a local structure similar to that of BPII.
Rheological behavior of BPIII presented very recently shows similar frequency dependence of $G'$ to that of the shear-induced phase in BPI~\cite{FSO2018}. 
Although the chemical components of BPI in this study are different from those of BPIII, the slow relaxation time might be a common feature in both systems. 
If we could prepare a stable BPIII over a wide temperature range~\cite{Iwa2010,Jakli2010} in the same liquid crystalline system, we could further clarify the different roles of ordered and disordered disclination networks and identify a structure of shear-induced phase in BPI rheologically.
Unfortunately, this is not possible with our chosen experimental system. 
\rev{Some liquid crystalline systems doped with achiral bent-core molecules show BPs over a wider temperature range without using polymer stabilization~\cite{Kikuchi2010,Jakli2010,Chien2017}. 
It would be possible to compare the rheological properties of the shear-induced phase with that of BPIII if we could use such an experimental system.}
%In fact, and to the best of our knowledge, there is currently no experimental system that features stability of several BPs over a wide temperature range without using polymer stabilization, something that precludes these systems from rheological studies.
Although further investigations are required to elucidate the structure of the shear-induced phase, defects 
play obviously an essential role for the rheological behavior of cubic BPI. 
In order to obtain detailed structural information, it is necessary to perform scattering experiments 
such as small angle X-ray scattering (SAXS).
For SAXS experiments impurities such as quantum dots or polymers had to be added to tag the disclination lines. Without impurities 
the electron density would be homogeneous throughout the sample and cannot lead to a meaningful scattering signal.
This hypothesis was confirmed though rheo-SAXS experiments at the Spring-8 Synchrotron, Japan.  
Kikuchi et al., for instance, performed SAXS measurement of blue phase by using a polymer tagged with iodine~\cite{Kikuchi15}. 
The inclusion of impurities, however, will lead to remarkable modification of rheological behavior, because adding them significantly stabilize the blue phase ~\cite{Yoshida09,Kara10,Draude20,Tang20,Guidimalla20}.
As impurities are very likely to fundamentally alter the viscoelastic properties of the pure blue phases ~\cite{Basappa}, introducing them is beyond the scope of this study. This important point will be addressed in future experiments.

\section{Acknowledgments}

OH acknowledges support from the EPSRC Early Career Fellowship Scheme (EP/N019180/2).
This work was supported by JSPS KAKENHI Grant Number 18H04469 and Ogasawara Toshiaki Memorial Foundation. 
\\

\appendix

\section{Flow curves and yield stress}

\begin{figure*}[htpb]
 \begin{center}
	\includegraphics[width=0.9\textwidth]{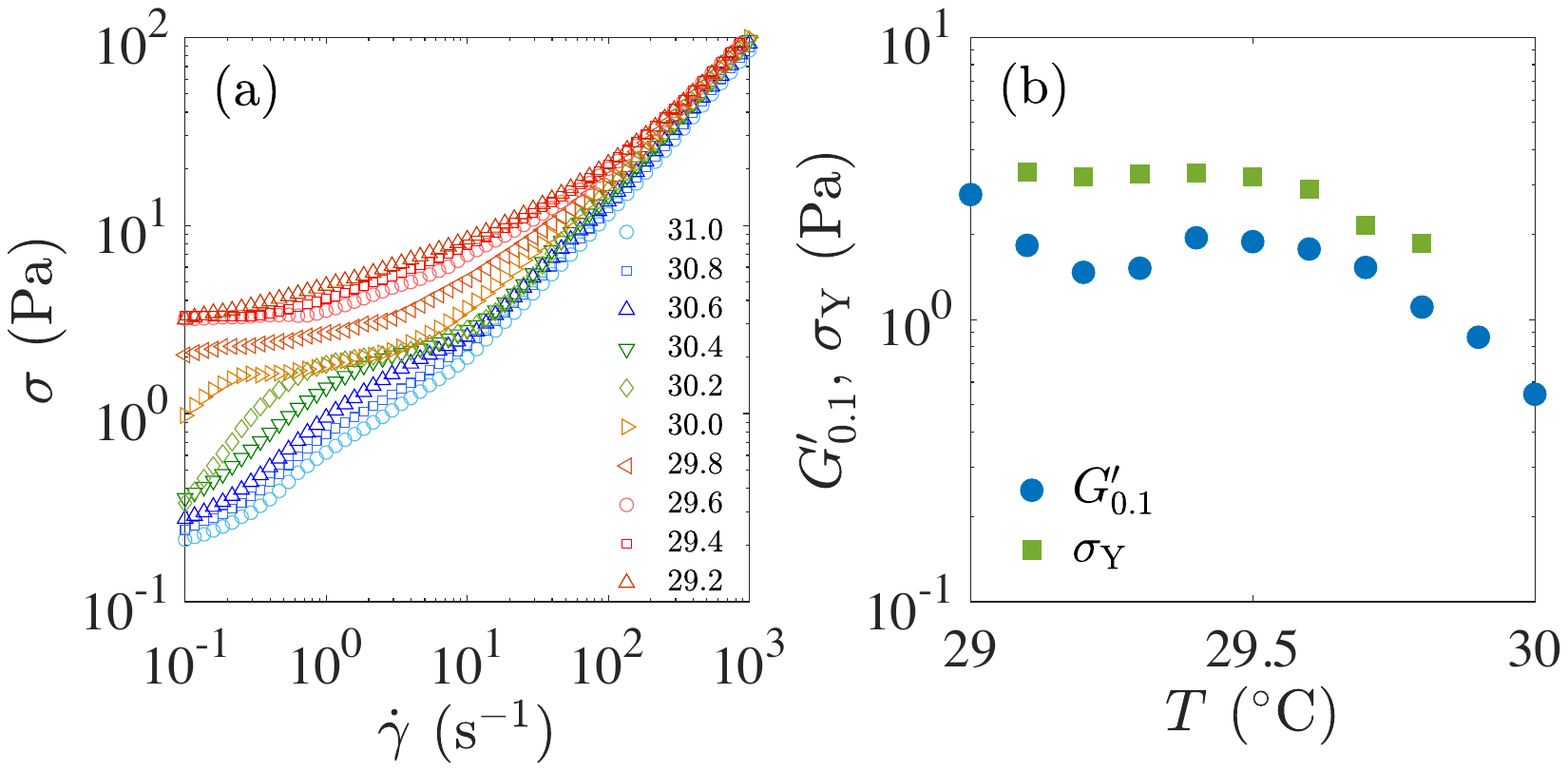}
 	\caption{
	(a) Flow curves at different temperatures. 
	The apparent yield stress can be estimated from the flow curves. 
	(b) Second plateau modulus $G'_{0.1}$ at $\omega$ = 0.1 s$^{-1}$ and apparent yield stress $\sigma_{\rm Y}$ as a function of temperature.
	}
	\label{figS1}
 \end{center}
\end{figure*}

Fig.~\ref{figS1}(a) provides flow curves at different temperatures measured by a strain-controlled rheometer ARES-G2. 
For yield stress measurements a stress-controlled rheometer is normally the more appropriate instrument as it allows to record the stress very accurately.
However, in our setup of the stress-controlled rheometer the shear cell is made of glass plates to facilitate the rheo-microscope observations of the textures. This leads often to inconsistencies in the rheological data between the measurements made with glass cells and normal stainless steel shear cells.  
To avoid such problems, we used the strain-controlled rheometer with the stainless steel shear cell for all measurements of the apparent yield stress.

In the flow curve measurements the subsequent sweep across shear rates was performed from high to low shear rates.
Thus, the samples were always pre-sheared.
Each shear rate was applied for 600 s and the viscosity was measured in the steady state and averaged over intervals of 120 s at every individual shear rate.
From the flow curves of the homogeneous BPI it is obvious that an apparent yield stress exists, whereas the BPI/Iso biphase does not exhibit an apparent yield stress. 

In Fig.~\ref{figS1}(b) we compare the second plateau modulus $G'_{0.1}$ at $\omega$ = 0.1 s$^{-1}$ from Fig. 5 of the main manuscript and the apparent yield stress $\sigma_{\rm Y}$.
It turns out the apparent yield stress is almost comparable to the second plateau modulus $G'_{0.1}$ measured without pre-shear. 
The correspondence of $\sigma_{\rm Y}$ to the second plateau modulus also confirms the viscoelastic solid-like properties of the homogeneous BPI as discussed in the main text.

\section{Supplemental movie}

Movie movieS1.mov is a supplementary movie to show the break-up of the platelet texture by shearing. 
The sample was sandwiched between two glass cover slips, while the bottom plate was kept fixed and the upper plate was moved by hand. 
Since the sample thickness is not controlled and the shear deformation was manually applied, we are unable to specify the exact shear rate. 
However, the experiment allows us to observe the process by which the platelet texture is broken up. 
After the shear deformation has been applied, we see a dark, homogeneous texture, which suggests a realignment of the disclination lines from an ordered to a disordered arrangement as the numerical simulation suggests. 
Although we can see the texture clearly, it is difficult to extract structural information even from this movie because we cannot control the shear rate and we do not see any visible textures after shearing.

\bibliography{BP1_Elasticity}

\end{document}